\documentclass[aps,pra,floats,floatfix,twocolumn,superscriptaddress,longbibliography]{revtex4-1}
\usepackage[utf8]{inputenc}
\usepackage{latexsym,amsmath}
\usepackage{amsbsy}
\usepackage[pdftex]{graphicx}
\usepackage{amssymb}
\usepackage{epstopdf}
\usepackage[usenames]{color}
\usepackage{float}
\usepackage{hyperref}
\usepackage{graphicx}
\usepackage{amsmath}
\usepackage[normalem]{ulem}
\usepackage{braket}

\newcommand{\sys}{\mathcal{S}}

\begin{document}

\title{Pairwise tomography networks for many-body quantum systems}

\author{Guillermo Garc\'ia-P\'erez}
\affiliation{QTF Centre of Excellence, Turku Centre for Quantum Physics, Department of Physics and Astronomy, University
of Turku, FI-20014 Turun Yliopisto, Finland}
\affiliation{Complex Systems Research Group, Department of Mathematics and Statistics,
University of Turku, FI-20014 Turun Yliopisto, Finland}

\author{Matteo A.~C.~Rossi}
\affiliation{QTF Centre of Excellence, Turku Centre for Quantum Physics, Department of Physics and Astronomy, University
of Turku, FI-20014 Turun Yliopisto, Finland}

\author{Boris Sokolov}
\affiliation{QTF Centre of Excellence, Turku Centre for Quantum Physics, Department of Physics and Astronomy, University
of Turku, FI-20014 Turun Yliopisto, Finland}

\author{Elsi-Mari Borrelli}
\affiliation{Corporate Research, ABB Switzerland Ltd., Baden, Switzerland}

\author{Sabrina Maniscalco}
\affiliation{QTF Centre of Excellence, Turku Centre for Quantum Physics, Department of Physics and Astronomy, University
of Turku, FI-20014 Turun Yliopisto, Finland}
\affiliation{QTF Centre of Excellence, Center for Quantum Engineering, Department of Applied Physics,
Aalto University School of Science, FIN-00076 Aalto, Finland}

\begin{abstract}
We introduce the concept of pairwise tomography networks to characterise quantum properties in many-body systems and demonstrate an efficient protocol to measure them experimentally. Pairwise tomography networks are generators of multiplex networks where each layer represents the graph of a relevant quantifier such as, e.g., concurrence, quantum discord, purity, quantum mutual information, or classical correlations. We propose a measurement scheme to perform two-qubit tomography of all pairs showing exponential improvement in the number of qubits $N$ with respect to previously existing methods. We illustrate the usefulness of our approach by means of several examples revealing its potential impact to quantum computation, communication and simulation. We perform a proof-of-principle experiment demonstrating pairwise tomography networks of $W$ states on IBM Q devices.
\end{abstract}

\maketitle

\section{Introduction}
The identification, characterisation and measurement of quantum properties in complex many-body systems is one of the greatest challenges of modern quantum physics. We are currently approaching a paradigm shift: on the one hand the quantum technology revolution is gaining speed as larger and more sophisticated quantum computers \cite{QComputingBook}, simulators \cite{QSimulationReview} and communication networks \cite{QCommsBook} are being developed and commercialised. On the other hand, we are acquiring increasing evidence of the presence and role of quantumness in real complex systems, such as biological \cite{QBiologyReview} and condensed matter systems \cite{QCondMattReview}.

We are now more than ever in need of novel multidisciplinary approaches, drawing on recent advances in complex network science, to tackle the formidable task of describing emergent and collective behaviour of quantum systems of increasing size and complexity \cite{BiamonteCNetworks}. The potential benefits of such a merging of disciplines would be remarkable. It would give us powerful tools to investigate questions such as: does quantumness play a functional role in biological systems? How can we optimise navigation and data transmission in the future quantum communication networks? How can we best engineer the quantum internet? How can we simulate complex new materials? How can we use quantum computers to design new chemical reactions and for drug discovery?

The results presented in this article lay the foundations of a novel interdisciplinary framework combining concepts of complex network science, quantum information, quantum measurement theory, and condensed matter physics. We introduce a new powerful tool able to capture, describe and visualise at once a class of quantum and classical properties in $N$-qubit systems, and we present an efficient measurement scheme to experimentally observe such properties. We focus on pairwise quantities, that is, those that can be computed from the two-qubit reduced density matrices (RDM) obtained by tracing out the remaining $N-2$ qubits.

Our main result is twofold. We firstly demonstrate in full generality how to perform pairwise tomography for all $N(N-1)/2$  pairs of qubits with only $\mathcal{O} \left( \log N \right)$ measurement settings. This constitutes an exponential improvement with respect to the expected scaling, which is polynomial in $N$.
Secondly, we introduce the concept of quantum tomography multiplexes, i.e., multilayer networks where the nodes are the qubits and, in every layer, the weighted links represent some (classical or quantum) pairwise quantity that can be directly obtained from the tomographic data. This results in a single mathematical object containing information about pairwise entanglement, mutual information, classical correlations, von Neumann entropy, quantum discord, or any other two-body quantifier which might be useful to characterise both many-body states and real quantum devices.

We illustrate the potential and usefulness of quantum tomography networks by considering several applications. While not containing the complete amount of information of the full $N$-qubit state, pairwise tomography networks allow us to investigate properties of many-body states and explore correlation in quantum critical systems \cite{Osterloh2002,Amico2008}, in particular in the context of quantum simulation \cite{friis2018}. Moreover, it has potential applications in quantum process tomography \cite{Govia2020}, in quantum chemistry \cite{McArdle2020} and in quantum computation (e.g., in SAT problems \cite{Kempe2006}, as well as in quantum machine learning \cite{Zhang2017}). Furthermore, as quantum technologies scale up, the investigation of complex states involving hundreds of qubits will be unfeasible unless a statistical perspective is taken, very much in the spirit of how the field of classical complex networks describes large complex structures.

The article is structured as follows. In Sec.~\ref{sec:measurement_scheme} we present the measurement scheme to reconstruct the pairwise tomography network and the corresponding multiplexes, and we prove that it scales logarithmically with $N$. In Sec.~\ref{sec:tomography_multiplex} we apply our method to reconstruct and characterise (i) the network of pairwise entanglement resulting from the experimental implementation of $W$ states on several IBM Q processors, (ii) system-environment states in an open quantum systems scenario, and (iii) the ground states of XX spin chains in a transverse magnetic field. Finally, in Sec.~\ref{sec:conclusion} we summarise our results and present conclusions and future perspectives.

\section{Measurement scheme for efficient pairwise tomography}
\label{sec:measurement_scheme}

The tomographic reconstruction of the quantum state of two qubits $i$ and $j$ requires the measurement of the nine correlators of the form $\langle \sigma_a^{(i)} \otimes \sigma_b^{(j)} \rangle$, where $\sigma_a$ and $\sigma_b$ represent Pauli matrices with $a$ and $b$ taking values $x$, $y$ and $z$. Therefore, characterising all pairwise density matrices in a system of $N$ qubits involves measuring $9 N(N-1)/2$ observables. However, if all qubits can be locally measured in any desired basis in every experimental realisation, it is possible to arrange the measurements in such a way that a much smaller number of measurement settings is needed. For instance, a simple parallelization scheme, in which one measures all non-overlapping pairs of qubits at once, reduces the number of measurement settings by a factor $\lfloor N /2 \rfloor$, thus bringing the number of required measurement setups to $\mathcal{O}(N)$.

In this section, we introduce a measurement scheme that allows us to obtain all these observables using only $\mathcal{O}\left( \log N \right)$ copies of the state. First, notice that all the correlators $\langle \sigma_x^{(i)} \otimes \sigma_x^{(j)} \rangle \, \forall i,j$ can be obtained via a single measurement setting in which all qubits are projected onto the $x$ basis, and similarly for the $y$ and $z$ bases. The correlators in which the two qubits are measured in different bases require more careful thinking.

Our measurement scheme relies on the assignment of three different labels, $a$, $b$, and $c$, to each qubit. These three labels are then taken to  represent measurement bases for each qubit, in such a way that any two different letters represent two different directions, $x$, $y$, or $z$. By letting these three letters run over all the six possible orderings of measurement bases, it is guaranteed that all the non-trivial correlators for any two qubits with different letters will be covered. However, no non-trivial correlators are measured for those pairs with equal letters. Hence, in this scheme, we aim at finding the minimal set of qubit labellings such that all pairs of qubits are covered.

Let us assume, without loss of generality, that each qubit is indexed by a different integer between $0$ and $N-1$. These integers can be represented in base three using only $\lceil \log_3 N \rceil$ digits, each of which can only take three different values. Our strategy is therefore simple: we use $\lceil \log_3 N \rceil$ labellings, indexed by $l = 1, \cdots, \lceil \log_3 N \rceil$, such that in labelling $l$ each qubit $i$ is assigned the letter $a$, $b$, or $c$, depending on the value of its $l$-th digit in the base-three representation of its index. Since any two different qubits have distinct indices, their base-three representation must have at least one different digit, so there will be at least one labelling in which their non-trivial correlators will be measured. Furthermore, it is clear that it is not possible to find a smaller number of labellings in which any two qubits are covered at least once; indeed, this would imply that one could assign a \textit{different} string of length $M \leq \lceil \log_3 N \rceil - 1$, each of them containing only letters $a$, $b$, and $c$, to each qubit. However, there are only $3^M \leq 3^{\lceil \log_3 N \rceil-1}$ such strings, while $N > 3^{\lceil \log_3 N \rceil-1}$.

Overall, the required number of different measurement settings is
\begin{equation}
6 \left\lceil \log_3 N \right\rceil + 3,
\end{equation}
that is, 6 settings per labelling plus the 3 trivial ones. This means that, for example, for around $N=50$ qubits, the size of the state-of-the-art NISQ devices available today, we need less than $30$ measurement settings, as opposed to more than $400$ settings needed with the na\"ive parallel approach.

Recently, similar algorithms for reducing the number of measurements required for $k$-wise tomography have been proposed \cite{yu2019quantum,cotler2019quantum,bonetmonroig2019nearly}. However, it should be stressed that, although our scheme is less general in that we only consider pairwise (as opposed to $k$-wise) tomography to construct the multiplex representations, it has a better scaling due to the fact that we label the qubits using three letters instead of two. As a consequence, we only require $6 \left\lceil \log_3 N \right\rceil + 3$, instead of $6 \left\lceil \log_2 N \right\rceil + 3$, measurement settings. In the example discussed in \cite{cotler2019quantum}, with $N=1024$ qubits, our algorithm requires $30\%$ less measurements.

\begin{figure*}[t]
    \centering
    \includegraphics[width = 0.9\linewidth]{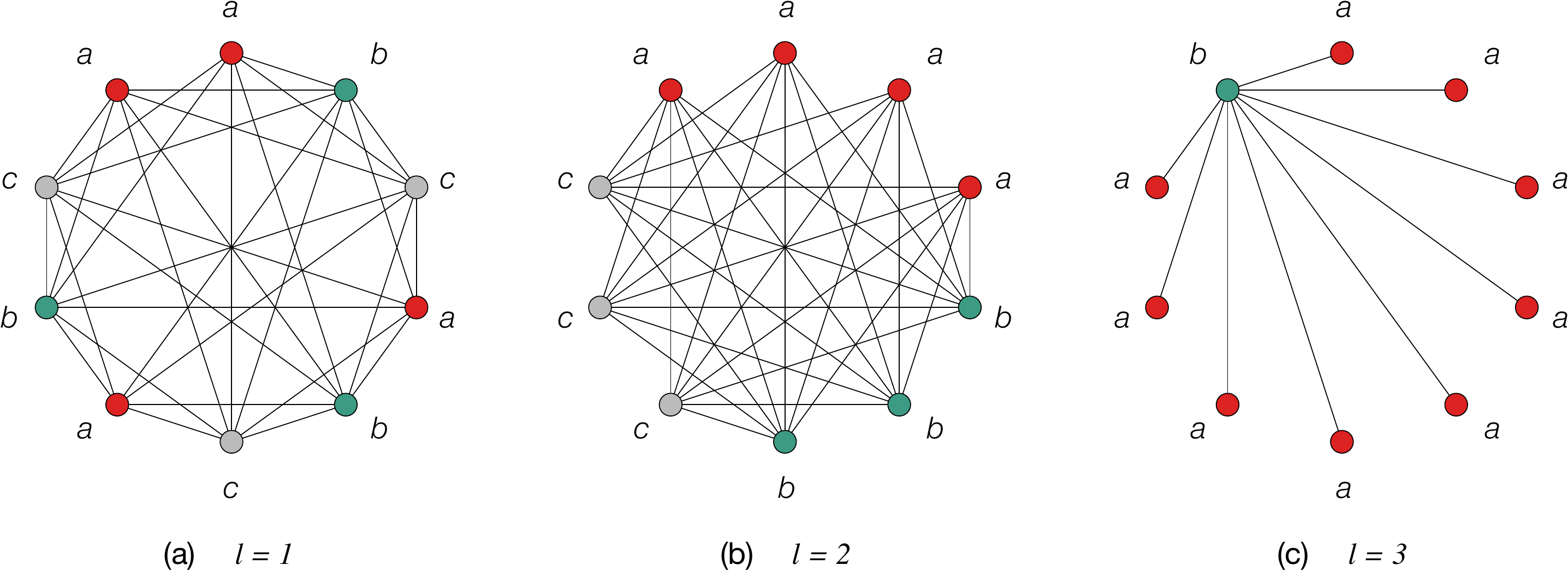}
    \caption{\label{fig:algorithm}Illustration of the steps 2.a in the measurement scheme for $N = 10$ qubits. Here,  $L = \lceil \log_3 10 \rceil=3$. Each figure depicts the letter assignment to each qubit (represented by three different colours: $a$ in red, $b$ in green, $c$ in gray) for $l = 1, 2, 3$ (from left to right). The connections represent the pairs of qubits for which the relevant observables are measured. Notice that the letters $a,b,c$ are assigned according to the result of $\left\lfloor i/3^{l-1} \right \rfloor \mod 3$, where $i$ is the qubit index (with $i = 0$ being the top-most qubit and the indices increasing in clockwise order).}
\end{figure*}

Operationally, our algorithm involves the following steps:

\begin{enumerate}
    \item Perform the three trivial measurements in which all qubits are measured along $x$, $y$, and $z$.
    \item Determine the number of different labellings needed: $ L = \lceil \log_3 N \rceil $. For $l=1,\ldots,L$, perform the following substeps:
    \begin{enumerate}
        \item Divide the qubits into groups of subsequent $3^{l-1}$ qubits and cyclically assign each group the letters $a,b,c,a,b,\ldots$. The last group may have less than $3^{l-1}$ qubits.
        \item Assign to $a, b, c$ all six permutations of $x, y, z$,
\begin{equation*}
\begin{matrix}
      &                & 1 & 2 & 3 & 4 & 5 & 6 \\
    a & \longleftarrow & x & x & y & y & z & z \\
    b & \longleftarrow & y & z & x & z & x & y \\
    c & \longleftarrow & z & y & z & x & y & x \\
\end{matrix}
\end{equation*}
and perform a measurement where each qubit is projected onto the direction indicated by the assigned letter.
\end{enumerate}
\end{enumerate}
Figure \ref{fig:algorithm} shows the $L=3$ different groupings of the qubits for the case $N=10$. For each $l$, subsequent groups of $3^{l-1}$ qubits are cyclically given the letters $a, b$ and $c$. The connections between qubits indicate that non-trivial correlations are measured between those qubits.

Once all the measurements in the scheme have been performed, we possess all the correlators required for the tomographic reconstruction of the RDM for each pair of qubits. The latter can be performed using any of the known methods such as simple linear inversion, maximum-likelihood \cite{Hradil1997,Banaszek1999,Rehacek2007,Smolin2012}, or Bayesian methods \cite{BlumeKohout2010}. We can thus form the so-called pairwise tomography network, in which every pair of qubits is assigned its corresponding RDM reconstructed from the tomographic data.
This network can then be unfolded into a \textit{quantum tomography multiplex} \cite{DeDomenico2013,Kivela2014,DeDomenico2015,BianconiMultiplexBook, cozzo_arruda_rodrigues_moreno_2018}, a multilayer network involving the qubits as nodes in which, in every layer, edges represent a different pairwise quantity.

In this work, we focus on six such quantities, namely mutual information, classical correlations, quantum discord \cite{Ollivier2001}, entanglement (measured via concurrence \cite{Wootters1998}), von Neumann entropy, and purity; to assign an edge between two qubits $i$ and $j$ in any of those layers, we simply compute the corresponding quantity from their RDM. For the classical correlations and quantum discord, non-symmetric quantities that depend on the choice of the measured qubit, we show the values obtained by performing the measurement on the qubit with the smallest index.

One should bear in mind  that, in general, it is possible to obtain non-zero values in correlation-related quantities as a consequence of mere fluctuations due to the finite amount of experimental data. However, in order to unveil the complex topological structure of the correlations of a given state, we filter out those links whose numerical value can be regarded as statistically irrelevant. To assess which connections are statistically significant, we apply a simple criterion: we first reconstruct the quantum tomography multiplex of a fully separable  pure state and, from it, we compute the mean and standard deviation of the weights in each layer for which these quantities should be null, e.g. concurrence, mutual information, etc. With these values, we can then consider the links whose value is larger than the mean plus five standard deviations in any other experiment as statistically significant.

\section{Quantum tomography networks: Applications}
\label{sec:tomography_multiplex}

In this section, we present applications of the pairwise tomography networks in the fields of quantum technologies and condensed matter physics. All the networks included in what follows have been obtained through a Qiskit \cite{Qiskit} implementation of our measurement scheme \cite{PairwiseTomographyGit},  either in simulated experiments using Qiskit's QASM Simulator or in real experiments on the freely available IBM Q Experience devices \texttt{ibmq\_burlington}, \texttt{ibmq\_essex}, \texttt{ibmq\_london}, \texttt{ibmq\_ourense}, and \texttt{ibmq\_vigo}. The tomographic reconstruction of the two-qubit density operator is done using Qiskit's tool, which employs a maximum-likelihood method proposed in Ref.~\cite{Smolin2012}.

\subsection{W states in IBM Q devices}
Here we consider, as an initial example of our measurement scheme, the  $W$ state, which belongs to a class of paradigmatic entangled states that have been extensively studied in the literature: the Dicke states.
Dicke states have gained widespread attention due to their usefulness in quantum metrology \cite{Dickemetro}, quantum game theory \cite{Dickegame}, quantum networks \cite{Dickenet}, and, interestingly, they have been proven useful for combinatorial optimization problems with hard constraints \cite{farhi2014,Hadfield2019}. They have been experimentally implemented in a variety of physical platforms, from trapped ions \cite{DickeExpIons1} to cold atoms \cite{DickeExpColdatoms1, DickeExpColdatoms2, DickeExpColdatoms3}, from superconducting qubits \cite{DickeExpSupercond1, DickeExpSupercond2} to photons \cite{DickeExpPhotons1, Dickenet}. Remarkably, Dicke states of over 200 qubits have been recently created in a solid-state platform \cite{Dicke200}.

An $N$-qubit $W$ state is defined as
\begin{equation}
    \vert W \rangle = \frac{1}{\sqrt{N}} \left( \vert 0\ldots01 \rangle + \vert 0\ldots10 \rangle + \cdots + \vert 1\ldots00 \rangle \right).
\end{equation}
The entanglement of this highly symmetric state is very robust against particle loss; indeed, the  state remains entangled even if any $N-2$ parties lose the information about their particle \cite{Dur2000,Dur2001}. This makes it particularly useful for quantum communication purposes as well as for robust quantum memories. In an $N$-qubit state, each pair of qubits possesses the same amount of entanglement with concurrence equal to $2/N$. Their corresponding entanglement network is therefore a fully connected graph. Moreover, $W$ states have the advantage of being efficiently implementable on gate-based quantum computers~\cite{Javerzac2019}. In particular, in Ref.~\cite{Javerzac2019}, the authors proposed an algorithm allowing to construct such states in very short times by parallelizing the required gates, and they tested it on an IBM Q processor (which is no longer available). In this work, we use their techniques to prepare $W$ states on five 5-qubit IBM Q devices as a proof-of-principle experiment for our pairwise tomography scheme.

\begin{figure}
    \includegraphics[width=.99\columnwidth]{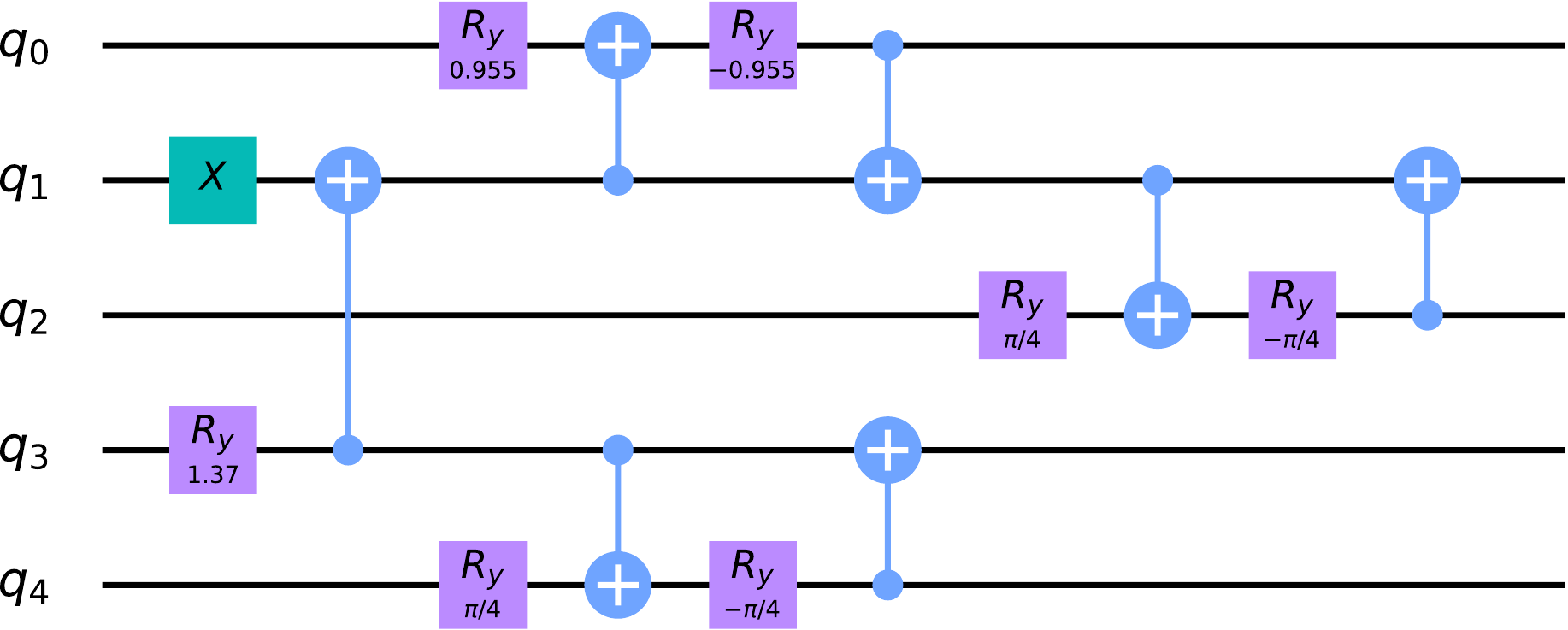}
    \caption{Circuit for the generation of a 5-qubit W state \cite{Javerzac2019} used in the experiments reported in Fig. \ref{fig:w_state}, optimized for the connectivity layout of the IBM Q devices used in the experiments, shown in Fig. \ref{fig:w_state}(f). The $R_y$ gates are rotations around the $y$ axis by the angle written below (in radians).}
    \label{fig:w_state_circuits}
\end{figure}

\begin{figure*}[t]
\centering
\includegraphics[width=.9\linewidth]{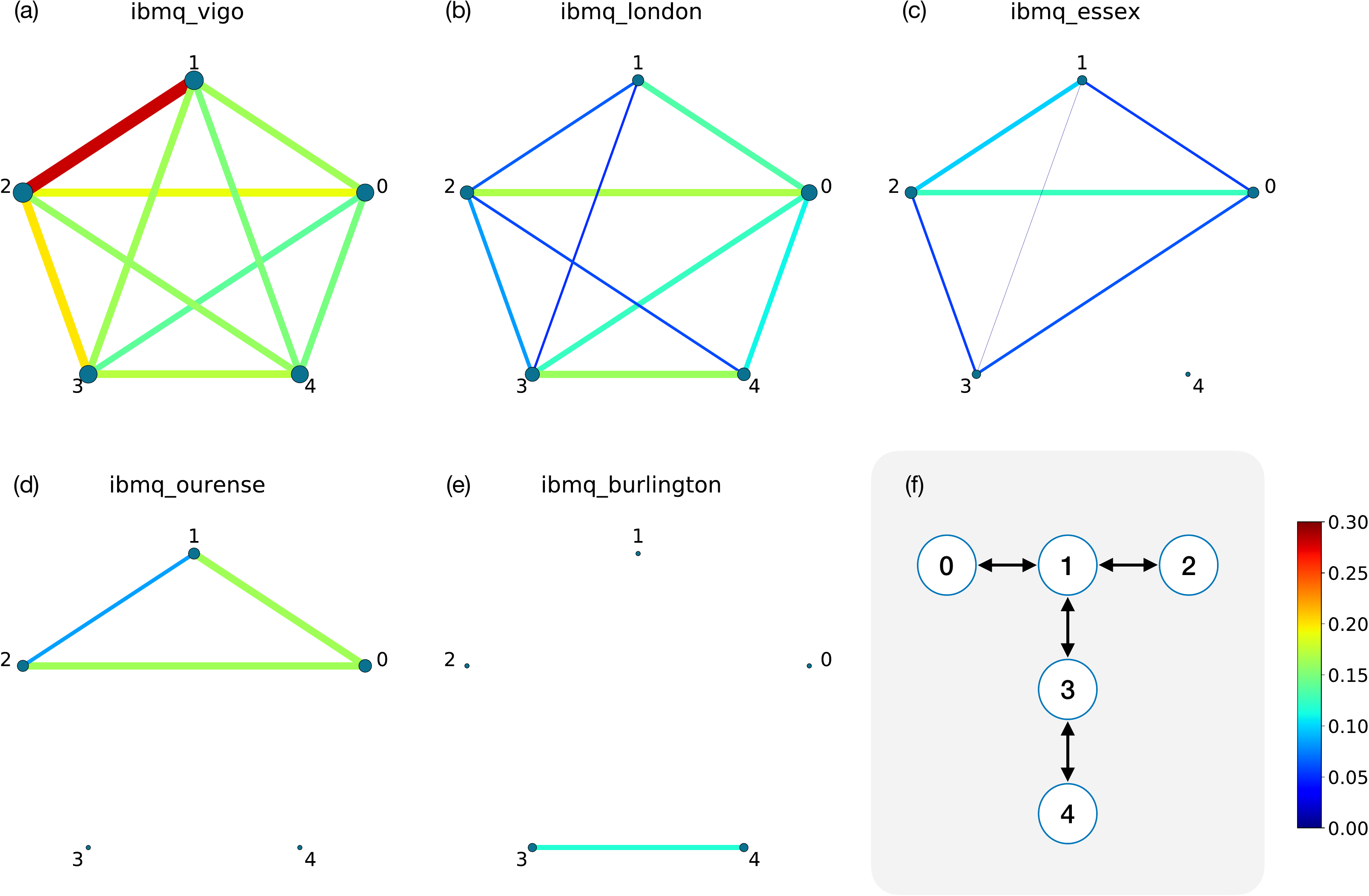}
\caption{(a)-(e) Pairwise concurrence network for a 5-qubit $W$ state generated experimentally on 5 different IBM Q devices. The edges represent the concurrence between the two qubits, represented as nodes, that it intersects. The weight of an edge is represented by its width and colour, which corresponds to the colour bar on the right. The size of each node is proportional to its strength, that is, the sum of the weights of the links reaching it. The edge between qubits 1 and 4 in \texttt{ibmq\_london}
has been filtered out, being statistically insignificant, as explained at the end of Sect.~\ref{sec:measurement_scheme}. For all other missing links, the reconstructed two-qubit states have zero concurrence. (f) the T-shaped connectivity layout of the IBM Q devices used in the experiment. The arrows represent the pairs of qubits that can be coupled with CNOT gates.}
\label{fig:w_state}
\end{figure*}

The working principle behind their algorithm is to first prepare an initial superposition of the form $\alpha |01\rangle + \beta |10\rangle$ between two qubits ($q_1$ and $q_3$ in this case) and then involve the rest of the qubits in the device by sequentially applying two-qubit gates that preserve the single-excitation subspace between connected pairs of qubits. The circuit used for the experiments in this work is shown in Fig. \ref{fig:w_state_circuits}; notice that two-qubit gates are only applied between qubits for which a CNOT gate can be physically implemented, in accordance with the connectivity layout of the devices (see Fig.~\ref{fig:w_state}(f)). While the circuit depicted in Fig. \ref{fig:w_state_circuits} is designed to prepare a 5-qubit $W$ state, additional single-qubit gates must be included in order to rotate the qubits for their measurement along the $x$ and $y$ directions when applying our pairwise tomography algorithm, since measurement in the IBMQ experience devices only allow measurements in the computational basis. In particular, these can be realized by an $H$ gate, or the combination of $S^\dagger$ and $H$ gates, respectively. In this work, we ran each measurement setting 8192 times (the maximum number of shots allowed) to gather statistics for the tomographic reconstruction.

\begin{table*}
\begin{tabular}{l@{\hspace{.5cm}}rrr@{\hspace{.8cm}}rr@{\hspace{.8cm}}r@{\hspace{.5cm}}r}
\toprule
Device &    $U_3$ &  CNOT &  Readout & $T_1$ ($\mu s$) &   $T_2$  ($\mu s$) & Q. volume & $\langle C \rangle$\\
\hline
\texttt{ibmq\_vigo}         &  0.26 \%  & 1.2  \%   & 2.2 \% &  76  &    65  & 16   & $0.17$ \\
\texttt{ibmq\_london}       &  0.075 \% & 1.2  \%   & 2.5 \% &  54  &    52  & 16   & $0.096$ \\
\texttt{ibmq\_essex}        &  0.099 \% & 1.5  \%   & 4.8 \% &  92  &   140  & 8  & $0.041$ \\
\texttt{ibmq\_ourense}      &  0.12 \%  & 0.96  \%  & 4.7 \% &  120 &    72  & 8    & $0.041$ \\
\texttt{ibmq\_burlington}   &  0.12 \%  & 1.6  \%   & 4.9 \% &  88  &    79  & 8    & $0.012$ \\
\toprule
\end{tabular}
\caption{Parameters of the IBM Q devices used in the experimental implementation of the $W$ state. From left to right: average error rate of single- ($U_3$) and two-qubit (CNOT) gates, average error rate of the readout, average $T_1$ and $T_2$ of the qubits, quantum volume \cite{quantumVolume}, and average pairwise concurrence $\langle C \rangle$  measured for the $W$ state. The error rates and decoherence times are obtained through the IBM Q API at the time the experiments were run.}
\label{tab:error_parameters}
\end{table*}

The results are shown in Fig.~\ref{fig:w_state}, where we present the networks of pairwise concurrence on the five devices. Each node corresponds to a different qubit in the device, and the thickness and color of each link correspond to the concurrence between the two qubits. As reported in Table~\ref{tab:error_parameters}, the IBM Q devices have significant error rates for single- and two-qubit gates, as well as in the measurement process. This results in missing links in the concurrence network, as well as a much lower average concurrence, compared to the expected one $\langle C \rangle = 2/5 = 0.4$. In addition, it is interesting to see that the correlation between the average errors reported in Table~\ref{tab:error_parameters} and the resulting average concurrence $\langle C \rangle$ is not straightforward, despite the connectivity layouts, as well as the circuit implementations of the $W$ state, being equal for all the processors. From an inspection of Table~\ref{tab:error_parameters}, the readout error appears to be the most impacting factor, followed by the error in CNOT gates, while single-qubit error, being around one order of magnitude smaller, is not significantly detrimental. The average concurrence does not seem to be correlated with the decoherence times $T_1$ and $T_2$, presumably because the depth of the circuit is relatively small.

We note a correlation between the quantum volume \cite{quantumVolume} of each device and the efficiency in the reconstruction of the expected pairwise tomography network, since \texttt{ibmq\_vigo} and \texttt{ibmq\_london} have higher quantum volume than all  the other devices. However, this simple quantifier of quantum computer power is not sufficient to grasp the rather notable difference in the performance of the two; the same is also true for \texttt{ibmq\_burlington} and \texttt{ibmq\_ourense}). Understanding the deviations from the theoretical prediction requires a more detailed device characterisation accounting for all the unwanted sources of error in the computers, which is beyond the scope of this paper.

\subsection{Decoherence in open quantum systems}

\begin{figure*}[t]
\centering
\includegraphics[width = 0.9\linewidth]{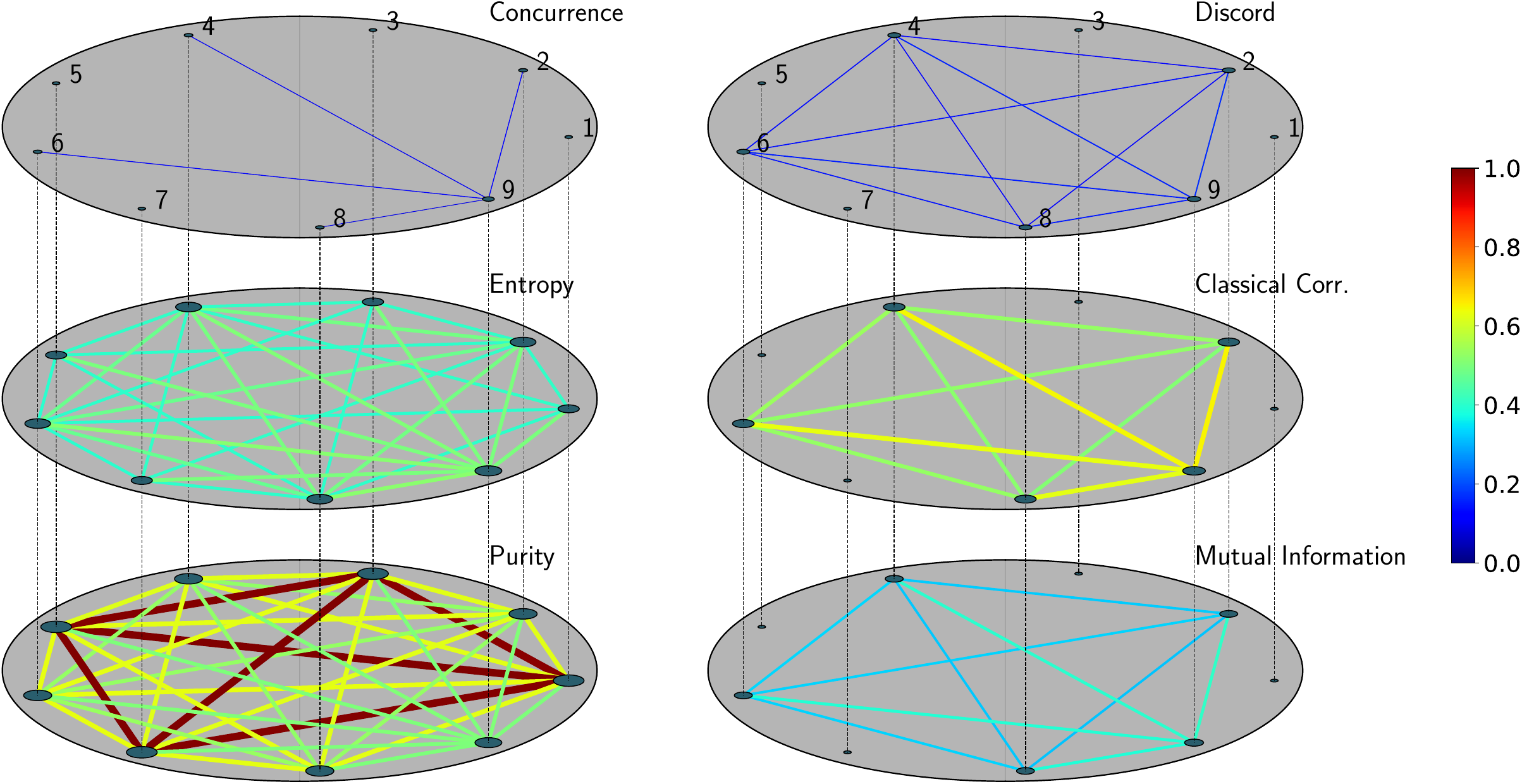}
\caption{Collisional model at time $\lambda t = 1000$, and with entangling interaction strength $\theta = 2 \pi / 3$. Qubits 1, 3, 5, and 7 are the emitters, 2, 4, 6, and 8, the ancillae, whereas 9 is the system qubit. The concurrence reveals pairwise entanglement between the system and the ancillae only, whereas there are quantum discord, classical correlations and mutual information among the ancillae as well. The emitters are in the ground state, since they are not correlated with any other qubit, while they form a clique in the purity layer; similarly, notice that there are no connections among emitters in the entropy layer.}
\label{fig:darwin}
\end{figure*}

The pairwise tomography network can also be used to generate multiplex representations of quantum states, in which the connections among qubits represent different quantifiers in every layer. This can be useful, for instance, for understanding the presence of correlations, quantum or classical, between an open quantum system and its quantum environment, as well as among the different parts of the latter. In order to illustrate this, we apply our machinery to the simulation of a collisional model in which a system qubit decoheres as a result of the interaction with other ancillary qubits (environment) at random times. In particular, we assume that each ancilla collides only once and at a time exponentially distributed with rate $\lambda/n$, where $n$ is the number of ancillae, and that the interaction between the system and an ancilla, driven by the Hamiltonian $H_I = \frac{\eta}{2} \sigma_x^a \otimes \sigma_z^\sys$, can be considered instantaneous, resulting in the unitary transformation $U_\theta = e^{-i \frac{\theta}{2} \sigma_x^a \otimes \sigma_z^\sys}$, where $\theta = \lim_{t \to 0} t \eta$ denotes the interaction strength and $t$ is the duration of the collision. Furthermore, we will consider the states of the system and an ancilla to be $| + \rangle_\sys$ and $| 0 \rangle_a$, respectively, before the collision.

It has been recently shown that this simple model can lead to the decoherence of the system \textit{even if the total state of system and ancillae remains fully separable at all times} as a consequence of the randomness in the collision times \cite{QDarwinism}. However, to illustrate the potential of the multiplex representation, we will consider entangling interactions in the current manuscript, as they lead to more complex quantum states.

We further give a quantum origin to the randomness in the collision times through the introduction of $n$ emitters, initially in the excited state, which relax to their ground state emitting an ancilla that immediately collides with the system qubit. Hence, if the initial state of the system is $| \psi_0 \rangle_\sys$, the total state for $n = 1$ at time $t$ is given by $\sqrt{e^{-\lambda t}} |1\rangle_e \otimes |0\rangle_a \otimes |\psi_0 \rangle_\sys + \sqrt{1 - e^{-\lambda t}} |0\rangle_e \otimes U_\theta \left(|0\rangle_a \otimes |\psi_0 \rangle_\sys \right)$; the generalisation for $n > 1$ is straightforward.

Although this dynamical process exhibits several interesting regimes as time evolves, we only focus on the long-time one here. However, we have created a video showing the time evolution of the multiplex, namely, how pairwise entanglement, quantum and classical correlations, and entropy/purity are dynamically established within the system-environment framework~\footnote{See the Supplemental Material at URL}. In Fig.~\ref{fig:darwin}, we show the multiplex of the corresponding state for $N = 9$ (that is, with 4 emitter-ancilla pairs), at time $\lambda t = 1000$, and with entangling interaction strength $\theta = 2 \pi / 3$. The resulting multiplex network exhibits a complex structure from which it is easy to identify the role of every qubit, i.e.~system, emitter, or ancilla, in the dynamics. The concurrence layer reveals that the system qubit is indeed entangled with all the ancillae but not with the emitters. Interestingly, despite the lack of entanglement between the different ancillae, these are nevertheless correlated, both at the classical and at the quantum levels, with non-zero classical correlations and discord (and, consequently, mutual information). Finally, the connectivity of the emitters reveals that, as expected at long times, they are in the ground state. This is consistent with the total lack of correlations with any other qubits and with the fact that the four emitters form a strongly connected clique in the purity layer; also, their connections are even deemed statistically irrelevant in the entropy layer.

\begin{figure*}[t]
\centering
\includegraphics[width = 0.9\linewidth]{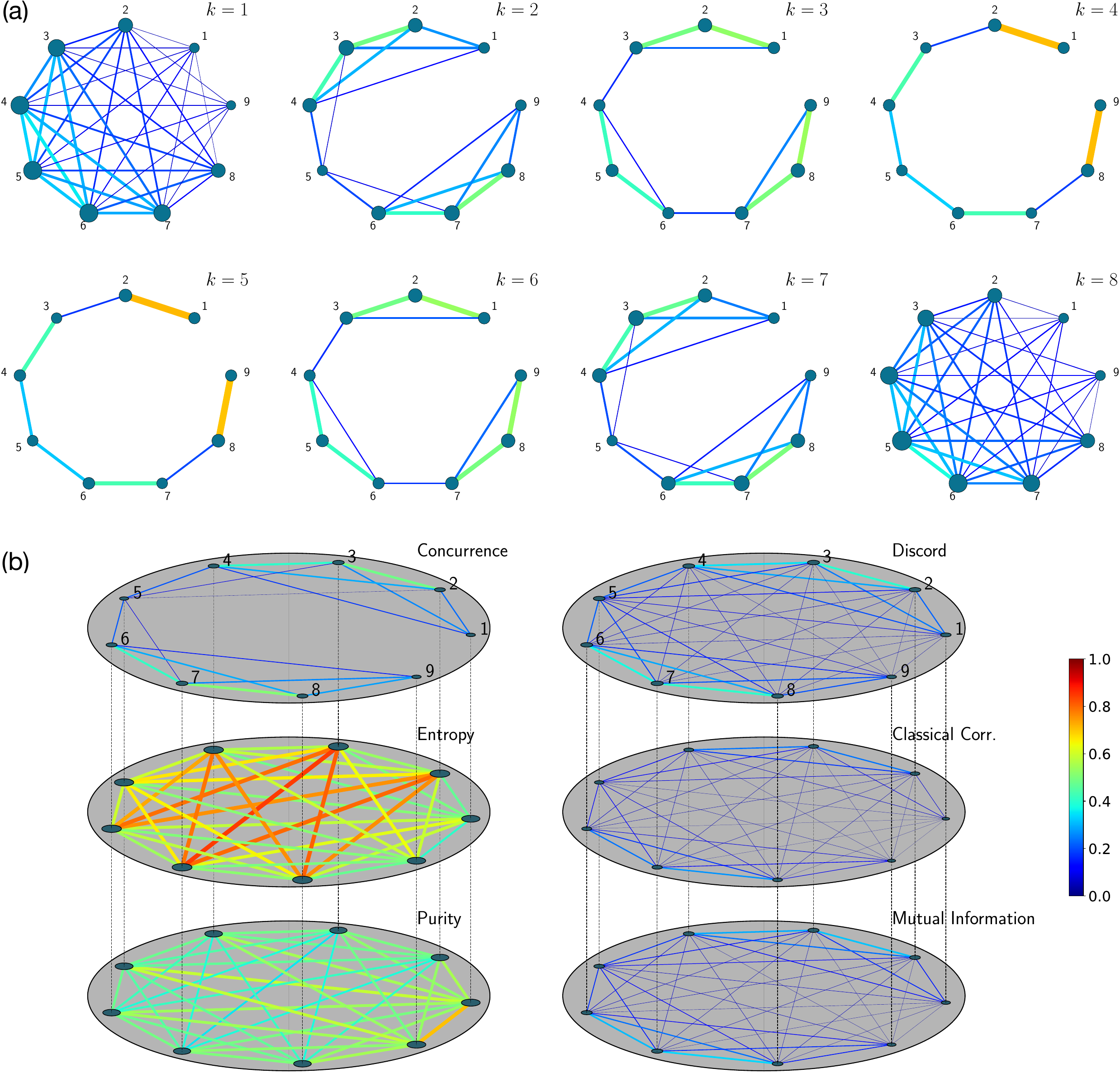}
\caption{(a) concurrence network of the XX ground state of an $N=9$ spin chain for different values of $k$. (b) pairwise multiplex for the ground state in the $k=2$ zone showing the differences in the two-spin properties between bulk pairs, e.g., spins 5 and 6, and edge pairs, e.g. spins 1 and 2.}
\label{fig:XX_concurrence}
\end{figure*}

\subsection{XX spin chain}
The full power of the pairwise tomography multiplex can be appreciated for systems displaying non trivial and non-homogeneous pairwise correlations. A perfect example is the spin-1/2  XX chain in a magnetic field, whose ground state possesses nontrivial topological order. In this model, the quasi-long-range order manifests itself in the formation of entangled edge states, with spins at the edge of the chain sharing entanglement quite differently from bulk spins \cite{QuantumInstability}.

The Hamiltonian of the spin chain is given by
\begin{equation}
    H=- J \sum_{i=1}^N \left[ \frac{1}{2}(\sigma^x_i \sigma^x_{i+1} + \sigma^y_i \sigma^y_{i+1}) + B \sigma^z_i \right],
\end{equation}
where $B$ is the magnetic field and $J$ is an overall coupling constant that we take as the unit of energy. In the following, we consider the open-boundaries case, i.e., we set $\sigma_{N+1} = 0$.

At finite size, the system is characterized by an instability of the ground state determined by a sequence of energy level crossings as the magnetic field is varied. This gives rise to sudden jumps in  pairwise entanglement which behaves non analytically. Following Ref.~\cite{QuantumInstability}, we indicate with $k$ the number of crossings, and with $B_{k+1}<B<B_k$ the corresponding regions of magnetic field, where $B_k=\cos[\pi k/(N+1)]$ are the critical values of $B$.

Pairwise entanglement jumps are very well captured by our concurrence networks (see Fig.~\ref{fig:XX_concurrence}), whose topology exhibits dramatic changes as we pass through different $k$ zones. The concurrence networks also clearly illustrate the difference in the entanglement of spin pairs in the bulk of the chain with respect to the edge. Specifically, one sees immediately the formation of entangled edge states (see, e.g., $k=5$), indicating the onset of long-range order in the system \cite{QuantumInstability}.

Let us now look at the pairwise tomography multiplex. Comparing the concurrence and entropy layers, we observe a somewhat expected and yet interesting phenomenon: there is a correlation between the weight of the edge connecting two qubits $i$ and $j$ in the entropy layer and their corresponding strengths $s_i$ and $s_j$ in the concurrence layer, where the strength of a node in a weighted network is defined as the sum of the weights of the edges intersecting it. This effect is especially visible by comparing the pairs $(3, 7)$ and $(1, 9)$. This correlation is a consequence of the fact that the pairwise state of two qubits that are highly entangled with other qubits is highly mixed (a similar anti-correlation can be observed between concurrence and the purity layers). Finally, the discord, classical correlations and mutual information graphs \cite{Valdez2017} are fully connected, showing that, even if pairwise entanglement is not present, other types of quantum and classical correlations are small but non-vanishing.

\section{Conclusions and Outlook}
\label{sec:conclusion}
We have introduced a new powerful concept for the characterization of  quantum and classical properties in many-body systems: the pairwise tomography networks. We have demonstrated that the they can be efficiently reconstructed experimentally and we presented a measurement scheme showing exponential improvement with respect to the known scaling. While containing less information than the full density matrix, pairwise RDM allow in-depth characterization of a quantum state with significantly less resources, with potential applications in quantum many-body physics \cite{Osterloh2002,Amico2008}, quantum chemistry \cite{McArdle2020}, quantum computation \cite{Kempe2006,Zhang2017}.

Applications of pairwise tomography networks to the investigation of quantum and classical properties of paradigmatic states, such as the $W$ states or the ground states of strongly correlated many-body systems, have been presented. These examples show that, through our new representation of two-body quantities, one may gain insight on the physical properties of complex quantum systems. Specifically, the network representation allows us to identify the distribution of pairwise quantum and classical properties within the many-body system. For stationary qubit systems, i.e., wherever the geometric location of the qubits is fixed, this may indicate the spatial distribution of quantum resources, such as entanglement, as exemplified by the XX spin chain model.

We have shown that further insight on the complex properties of many-body systems can be obtained by means of multiplex networks, where different pairwise properties are represented in different layers. Topological correlations between the different network layers may reveal additional information about the structure of the underlying state. Another possible interesting application of multiplex networks, which we plan to study in the future, is the case in which the system evolves dynamically, as in the collisional model here described. Each layer of the multiplex network may represent, e.g., pairwise entanglement at different times, allowing us to study the temporal correlations in the dynamical evolution of such quantity.

Multiplex networks are extensively studied in network science, and there exist tools for analyzing statistically their properties.
While the examples considered in this papers are all meant to illustrate, as proof-of-principle, the potential of these concepts, we envisage several scenarios in which, for increasing $N$, statistical methods from classical network theory will be needed to characterise the system's properties.
In this sense our results may stimulate a much sought cross-fertilization between complex network science and quantum many-body physics. This, in turn, may be a key ingredient for the emergence of a new approach to answer both fundamental and applicative questions in quantum biology, quantum chemistry, quantum technologies, and condensed matter physics.

\begin{acknowledgments}
We acknowledge the use of the IBM Q Experience for this work. The views expressed are those of the authors and do not reflect the official policy or position of IBM or the IBM Q team. G.G.-P., M.A.C.R., B.S. and S.M.~acknowledge financial support from the Academy of Finland via the Centre of Excellence program (Project no.~312058). G.G.-P.~acknowledges support from the emmy.network foundation under the aegis of the Fondation de Luxembourg. B.S. acknowledges financial support from the Jenny and Antti Wihuri Foundation.
\end{acknowledgments}

\bibliography{bibliography}

\end{document}